# A note on proper affine symmetry in Bianchi types VIII and IX space-times

Ghulam Shabbir

Faculty of Engineering Sciences,

GIK Institute of Engineering Sciences and Technology,

Topi, Swabi, NWFP, Pakistan,

Email: shabbir@giki.edu.pk

and

M. Rafiq

Centre for advanced studies in Engg. (CASE) Islamabad, Pakistan.

**Abstract**

We consider the most general form of Bianchi types VIII and IX space-times for studying proper affine symmetry by using holonomy and decomposability, the rank of the $6\times 6$ Riemann matrix and direct integration techniques. Studying proper affine symmetry in the above space-times it is shown that there exists only one possibilty when the above space-times admit proper affine vector field.

## 1. INTRODUCTION

Affine symmetry which preserves the geodesic structure and affine parameter of a space-time carries significant information and interest in the Einstein's theory of general relativity. It is therefor important to study this symmetry. In this paper we are interested to find the existence of proper affine vector fields in Bianchi types VIII and IX space-times by using holonomy and decomposability, the rank of the $6\times 6$ Riemann matrix and direct integration techinques. Let $(M,g)$ be a space-time with $M$ a smooth connected Hausdorff four dimensional manifold and $g$ a smooth metric of Lorentz signature (-, +,

+, +) on $M$. The curvature tensor associated with $g$, through Levi-Civita connection, is denoted in component form by $R^a{}_{bcd}$. The usual covariant, partial and Lie derivatives are denoted by a semicolon, a comma and the symbol $L$, respectively. Round and square brackets denote the usual symmetrization and skew-symmetrization, respectively. The space-time $M$ will be assumed nonflat in the sense that the Riemann tensor does not vanish over any non-empty open subset of $M$.

A vector field $Z$ on $M$ is called an affine vector field if it satisfies

$$Z_{a;bc} = R_{abcd} Z^d \tag{1}$$

or equivalently,

$$Z_{a,bc} - \Gamma^f_{ac} Z_{f,b} - \Gamma^f_{bc} Z_{a,f} - \Gamma^e_{ab} Z_{e,c} + \Gamma^e_{ab} \Gamma^f_{ec} Z_f - (\Gamma^e_{ab})_{,c} Z_e - \Gamma^f_{ab} \Gamma^e_{cf} Z_e$$
$$+ \Gamma^e_{fb} \Gamma^f_{ca} Z_e + \Gamma^e_{af} \Gamma^f_{bc} Z_e = R_{abcd} Z^d.$$

The covariant derivative of any vector fields $Z$ on $M$ can be decomposed into its symmetric and skew-symmetric parts

$$Z_{a;b} = \frac{1}{2} h_{ab} + F_{ab} \qquad (h_{ab} = h_{ba}, \ F_{ab} = -F_{ba}) \tag{2}$$

then equation (1) is equivalent to

$$(i) \ h_{ab;c} = 0 \quad (ii) \ F_{ab;c} = R_{abcd} Z^d \quad (iii) \ F_{ab;c} Z^c = 0. \tag{3}$$

If $h_{ab} = 2c g_{ab}, c \in R$, then the vector field $Z$ is called homothetic (and *Killing* if $c = 0$). The vector field $Z$ is said to be proper affine if it is not homothetic vector field and also $Z$ is said to be proper homothetic vector field if it is not Killing vector field on $M$ [3]. Define the subspace $S_p$ of the tangent space $T_p M$ to $M$ at $p$ as those $k \in T_p M$ satisfying

$$R_{abcd} k^d = 0. \tag{4}$$

## 2. Affine Vector Fields

Suppose that $M$ is a simple connected space-time. Then the holonomy group of $M$ is a connected Lie subgroup of the idenity component of the Lorentz group and is thus characterized by its subalgebra in the Lorentz algebra. These have been labeled into fifteen types $R_1 - R_{15}$ [2]. It follows from [3] that the only such space-times which could admit proper affine vector fields are those which admit nowhere zero covariantly constant second order symmetric tensor field $h_{ab}$. This forces the holonomy type to be either $R_2$, $R_3$, $R_4$, $R_6$, $R_7$, $R_8$, $R_{10}$, $R_{11}$ or $R_{13}$ [3]. A study of the affine vector fields for the above holonomy type can be found in [3]. It follows from [3] that the rank of the $6\times 6$ Riemann matrix of the above space-times which have holonomy type $R_2$, $R_3$, $R_4$, $R_6$, $R_7$, $R_8$, $R_{10}$, $R_{11}$ or $R_{13}$ is atmost three. Hence for studying affine vector fields we are interested in those cases when the rank of the $6\times 6$ Riemann matrix is less than or equal to three.

## 3. MAIN RESULTS

Consider the space-times in the usual coordinate system $(t,x,y,z)$ (labeled by $(x^0,x^1,x^2,x^3)$, respectively) with line element [1]

$$ds^2 = -dt^2 + A(t)(dx + f'(y)dz)^2 + B(t)dy^2 + B(t)f^2(y)dz^2, \qquad (5)$$

where $A(t)$ and $B(t)$ are no where zero functions of $t$ only and prime denotes the derivative with respect to $y$. For $f(y)=\sin y$ or $f(y)=\sinh y$ the above space-time (5) becomes Bianchi type IX or Bianchi type VIII space-time, respectively. The above space-time admits four linearly independent Killing vector fields which are

$$\frac{\sin z}{f(y)}\frac{\partial}{\partial x}+\cos z\frac{\partial}{\partial y}-\frac{f'(y)}{f(y)}\sin z\frac{\partial}{\partial z},\ \frac{\partial}{\partial x},\ \frac{\partial}{\partial z},\ -\frac{\cos z}{f(y)}\frac{\partial}{\partial x}+\sin z\frac{\partial}{\partial y}+\frac{f'(y)}{f(y)}\cos z\frac{\partial}{\partial z}. \qquad (6)$$

The non-zero independent components of the Riemann tensor are

$$R^0{}_{101} = -\frac{1}{4}(\frac{A_t^2-2AA_{tt}}{A}),\ R^0{}_{202} = -\frac{1}{4}(\frac{B_t^2-2BB_{tt}}{B}),$$

$$R^0{}_{303} = -\frac{1}{4}[(\frac{A_t^2-2AA_{tt}}{A})f'^2(y)+(\frac{B_t^2-2BB_{tt}}{B})f^2(y)],$$

$$R^0{}_{123} = -(\frac{A_tB-AB_t}{2B})f(y),\ R^0{}_{103} = -\frac{1}{4}(\frac{A_t^2-2AA_{tt}}{A})f'(y),$$

$$R^1{}_{212} = (\frac{A^2+BA_tB_t}{4AB}),\ R^0{}_{323} = 3(\frac{AB_t-BA_t}{4B})f(y)f'(y), \qquad (7)$$

$$R^1{}_{313} = \frac{1}{4}[(\frac{A_tB_tB+A^2-AB}{B^2})f'^2(y)+\frac{A_tB_t}{A}f^2(y)+\frac{A}{B}],$$

$$R^2{}_{323} = \frac{1}{4}[(\frac{A_tB_tB+A^2-B^2}{B^2})f'^2(y)+\frac{B_t^2}{B}f^2(y)+\frac{4B-3A}{B}],$$

$$R^1{}_{232} = \frac{1}{4}[\frac{A_tB_tB+4A^2-4AB-AB_t^2}{AB}]f'(y).$$

Writing the curvature tensor with components $R_{abcd}$ at $p$ as a $6\times 6$ symmetric matrix [4]

$$R_{abcd} = \begin{pmatrix} R_{0101} & 0 & R_{0103} & 0 & 0 & R_{0123} \\ 0 & R_{0202} & 0 & 0 & 0 & 0 \\ R_{0301} & 0 & R_{0303} & 0 & 0 & R_{0323} \\ 0 & 0 & 0 & R_{1212} & 0 & R_{1223} \\ 0 & 0 & 0 & 0 & R_{1313} & 0 \\ R_{2301} & 0 & R_{2303} & R_{2312} & 0 & R_{2323} \end{pmatrix}.$$

As mentioned in section 2, the space-times which can admit proper affine vector fields have holonomy type $R_2$, $R_3$, $R_4$, $R_6$, $R_7$, $R_8$, $R_{10}$, $R_{11}$ or $R_{13}$ and the rank of the $6\times 6$ Riemann matrix is atmost three. Hence, we are only interested in those cases when the rank of the $6\times 6$ Riemann matrix is less than or equal to three (excluding the flat cases), thus after some lengthy and tedious calculation we find that there exists only one possibility which is: $A_t = 0$, $B_t = 0$ and the rank of the $6\times 6$ Riemann matrix is 3.

In this case $A_t = 0$, $B_t = 0$, the rank of the $6\times 6$ Riemann matrix is 3 and there exists a unique (up to a multiple) nowhere zero timelike vector field $t_a = t_{,a}$ satisfying $t_{a;b} = 0$ (and from the Ricci identity $R^a{}_{bcd} t_a = 0$). Equations $A_t = 0$ and $B_t = 0 \Rightarrow A = a$ and $B = b$, where $a, b \in R \setminus \{0\}$. Here, we assume that $a \neq b$. The line element in this case is

$$ds^2 = -dt^2 + a(dx + f'(y)dz)^2 + b\, dy^2 + b f^2(y) dz^2, \tag{8}$$

The above space-time is 1+3 decomposable. Affine vector fields in this case [3] are

$$Z = (c_5 t + c_6) \frac{\partial}{\partial t} + Z', \tag{9}$$

where $c_5, c_6 \in R$ and $Z'$ is a homothetic vector field in the induced geometry on each of the three dimensional submanifolds of constant $t$. The completion of the case needs to find a homothetic vector field in the induced geometry of the submanifolds of constant $t$. The induced metric $g_{\alpha\beta}$ (where $\alpha, \beta = 1, 2, 3$) with non zero components is given by

$$g_{11} = a,\ g_{22} = b,\ g_{33} = a f'^2(y) + b f^2(y),\ g_{13} = a f'(y). \tag{10}$$

A vector field $Z'$ is a homothetic vector field if it satisfies $L_{Z'} g_{\alpha\beta} = 2\phi g_{\alpha\beta}, \forall a,b = 1,2,3,$ where $\phi \in R$. One can expand the homothetic equation and using (10) to get

$$Z^1_{,1} + f'(y)Z^3_{,1} = \phi \tag{11}$$

$$a Z^1_{,2} + b Z^2_{,1} + a f'(y)Z^3_{,2} = 0 \tag{12}$$

$$a f''(y)Z^2 + a Z^1_{,3} + (a f'^2(y) + b f^2(y))Z^3_{,1} + a f'(y)(Z^1_{,1} + Z^3_{,3}) = 2\phi a f'(y) \tag{13}$$

$$Z^2_{,2} = \phi \tag{14}$$

$$b Z^2_{,3} + (a f'^2(y) + b f^2(y))Z^3_{,2} + a f'(y)Z^1_{,2} = 0 \tag{15}$$

$$2\{a f''(y) + b f(y)\}f'(y)Z^2 + 2a f'(y)Z^1_{,3} + 2(a f'^2(y) + b f^2(y))Z^3_{,3} = 2\phi(a f'^2(y) + b f^2(y)). \tag{16}$$

Equation (14) implies $Z^2 = \phi y + F^1(x,z)$, where $F^1(x,z)$ is a function of integration. Differentiating equation (11) and (12) with respect to $y$ and $x$, respectively and after subtracting we get $Z^3_{,1} = \dfrac{b}{af''(y)} Z^2_{,11}$. Substituting the value of $Z^2$ in the previous equation and upon integration we get

$$Z^3 = \dfrac{b}{af''(y)} F^1_x(x,z) + F^2(y,z), \tag{17}$$

where $F^2(y,z)$ is a function of integration. Using the above information in equation (11) we get $Z^1 = \phi x - \dfrac{bf'(y)}{af''(y)} F^1_x(x,z) + F^3(y,z)$, where $F^3(y,z)$ is a function of integration.

One has

$$\begin{aligned} Z^1 &= \phi x - \dfrac{bf'(y)}{af''(y)} F^1_x(x,z) + F^3(y,z), \\ Z^2 &= \phi y + F^1(x,z), \\ Z^3 &= \dfrac{b}{af''(y)} F^1_x(x,z) + F^2(y,z). \end{aligned} \tag{18}$$

In order to determine $F^1(x,z)$, $F^2(y,z)$ and $F^3(y,z)$ we need to integrate remaining three equations. To avoid lengthy calculation we will only present result here. If one proceeds further, after some calculation one finds that $\phi = 0 \Rightarrow$ no proper homothetic vector field exist in the induced geometry of the submanifolds of constant $t$. Hence the homothetic vector fields in this case are Killing vector fields which are

$$Z^1 = \frac{\sin z}{f(y)} c_1 - \frac{\cos z}{f(y)} c_2 + c_3,$$
$$Z^2 = \cos z\, c_1 + \sin z\, c_2, \tag{19}$$
$$Z^3 = -\frac{f'(y)}{f(y)} \sin z\, c_1 + \frac{f'(y)}{f(y)} \cos z\, c_2 + c_4,$$

where $c_1, c_2, c_3, c_4 \in R$. Affine vector fields in this case are (from equation (9) and (19))

$$Z^0 = (c_5 t + c_6), \qquad Z^1 = \frac{\sin z}{f(y)} c_1 - \frac{\cos z}{f(y)} c_2 + c_3,$$
$$Z^2 = \cos z\, c_1 + \sin z\, c_2, \tag{20}$$
$$Z^3 = -\frac{f'(y)}{f(y)} \sin z\, c_1 + \frac{f'(y)}{f(y)} \cos z\, c_2 + c_4.$$

One can write the above equation (20) after subtracting Killing vector fields as

$$Z = (t, 0, 0, 0) \tag{21}$$

Clearly, in this case the above space-time (8) admits proper affine vector field.

## SUMMARY

In this paper a study of Bianchi types VIII and IX space-times according to their proper affine vector fields is given. An approach is developed to study proper affine vector fields in the above space-times by using the rank of the $6 \times 6$ Riemann matrix and holonomy. From the above study we obtain that there exists only one possibilty when the above space-times admit proper affine vector fields. This is the case when the rank of the $6 \times 6$ Riemann matrix is three and there exists a nowhere zero independent timelike vector field which is the solution of equation (4) and is covariantly constant. In this case the space-times are given in equation (8) and they admit proper affine vector fields (for details see equation (20)).